# Exploring the experimental foundation with rupture and delayed rupture

Asal Siavoshani, Ming-Chi Wang, Cheng Liang, Aanchal Jaisingh[#], Junpeng Wang,
Chen Wang[#], Shi-Qing Wang[*]

School of Polymer Science and Polymer Engineering
University of Akron, Akron, Ohio 44325
#: Department of Materials Science and Engineering
University of Utah, Salt Lake City, UT 84112

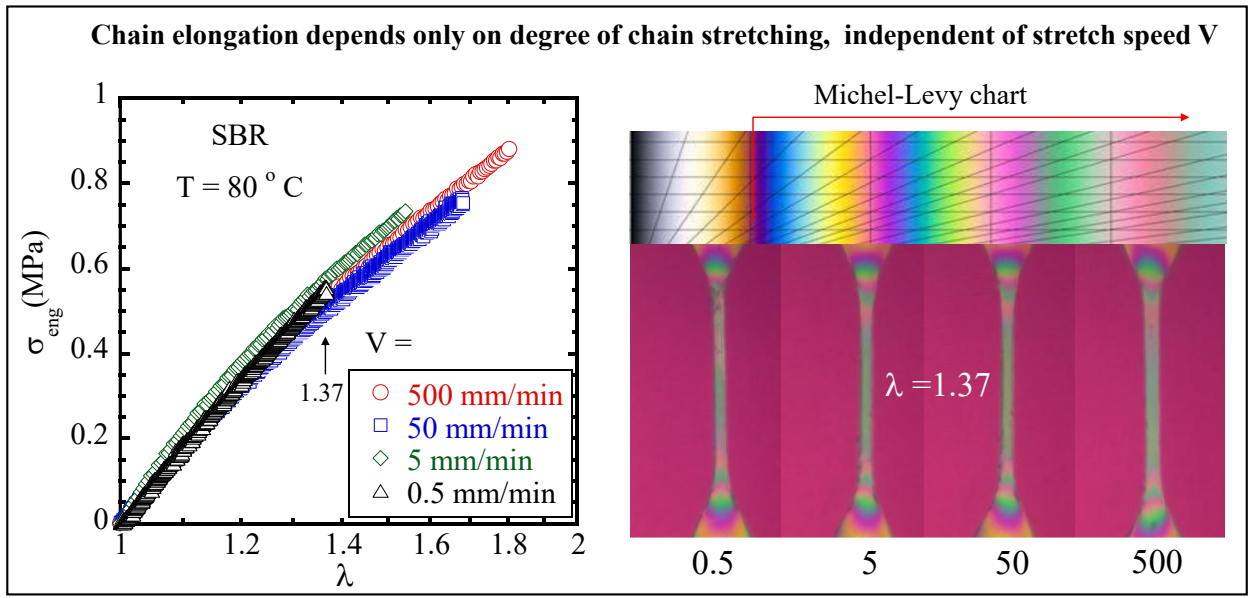

For Table of Contents use only

## Abstract

We carry out uniaxial continuous and step stretching of various crosslinked polymer networks to demonstrate how characteristics of rupture (from continuous stretching) and delayed rupture (from step stretching) can be used to probe the structure of the emergent kinetic theory of bond dissociation (KTBD) for elastomeric failure. Based on delayed rupture experiments, we show that the network lifetime $t_{ntw}$, taken as the incubation time $t_{del-rupt}$ for delayed rupture, depends on temperature in an Arrhenius like manner and is exponentially sensitive to the degree of network stretching (depicted by step-stretch ratio $\lambda_{ss}$). Rupture during continuous stretching for a wide range of stretch rates takes place on timescales inversely proportional to the stretch rate. The elapsed time $t_{rupt}$ at rupture is found to be comparable to $t_{del-rupt}$ at various values of $\lambda_b = \lambda_{ss}$ in a wide range of temperature, affording the experimental basis for the premise of the KTBD. Having identified the hidden internal clock $t_{ntw}$, continuous stretching tests at different temperatures are performed to show the existence of a new time temperature equivalence (TTE): fast stretching at higher temperatures is equivalent to slow stretching at lower temperatures: different pairs of rate and temperature can produce the rupture at the same tensile strength and strain.

---

[*] Corresponding authors at swang@uakron.edu.

1. Introduction

Elastomeric rupture, resulting from continuous stretching of *unnotched* specimens until sudden macroscopic separation, has been treated in literature as a fracture mechanical phenomenon by assuming the preexistence of sizable flaws. Consequently, instead of an extensive illustration of rupture characteristics following the pioneering investigations[1-3] of Smith and others, nearly all studies in the past[4-7] have focused on fracture behavior of *prenotched* systems. Recent spatially temporally resolved polarized optical microscopic (*str*-POM) measurements of several crosslinked rubbery polymers showed[8,9] that the tip birefringence upon visible crack propagation is comparable to that observed at rupture. Therefore, elastomers may be generally flaw insensitive, and rupture may be treated and understood outside the scope of fracture mechanics. This realization made it possible to develop new theoretical understanding of temperature effect on elastomeric failure.

Temperature controls mechanical behavior of polymeric materials in several crucial ways. In polymer processing, there is the well-known time-temperature equivalence (TTE$_{ve}$) associated with viscoelasticity,[10] dictating rheology of polymers in liquid (molten or rubbery) states.[11,12] Here, Williams, Landel and Ferry (WLF) found[10] polymer viscosity and chain relaxation time $\tau_{ve}$ to explicitly vary with temperature. It has since been well-established that slower relaxation dynamics at lower temperatures are associated with higher viscosity. For seven decades,[13,14] researchers advocated[15-20] that the same viscoelastic processes dictated time and temperature dependencies of rupture and fracture in elastomers. This viewpoint has remained unchallenged until recently,[21] even though WLF shift did not produce[22] any clear explanation for the observed temperature dependence of tensile strength from rupture and energy release rate for crack growth.

A different time-temperature equivalence (TTE$_{ntw}$), hidden for seven decades, has recently been suggested[21] to characterize elastomeric rupture and fracture, where the subscript "ntw" stands for the network lifetime $t_{ntw}$ associated with bond dissociation in highly stretched elastomers. The lifetime $t_{ntw}$ of the chain network is the hidden internal clock in elastomers. A network, composed of permanent chemical crosslinks and transient physical crosslinks due to interchain uncrossability,[23] can disintegrate when a percolating fraction of load bearing strands (LBS) undergoes chain scission. Determined by bond dissociation in backbones of LBS in high tension produced by stretching at (nominal) stretch ratio $\lambda$, $t_{ntw}$ depends on temperature T in an Arrhenius manner, e.g., showing an exponential dependence on reciprocal temperature 1/T. Moreover, we expect $t_{ntw}(\lambda, T)$ to vary with the network structure that may be crudely characterized in terms of crosslink density.

Built on earlier treatments that regard overcoming solid strength as an activated process,[24, 25] the recent kinetic theory of bond dissociation (KTBD) for rupture[21, 26] asserts that rupture occurs because the elapsed time till rupture, given by

$$t_{rupt} = (\lambda_b - 1)/\dot{\lambda}, \tag{1}$$

approaches $t_{ntw}$, as illustrated in Figure 1. In other words, continuous stretching of an unnotched elastomer till rupture produces a measure of the network lifetime $t_{ntw}$. Here, for simplicity, we express $t_{ntw}$ in an Arrhenius-like form

$$t_{ntw}(T, \lambda) = t_{ntw0}(T)\exp[E_{ntw}(\lambda)/RT], \tag{2}$$

which also rapidly decreases with increasing $\lambda$ since the energy barrier $E_{ntw}$ for network breakdown is a decreasing function of the level of network stretching.

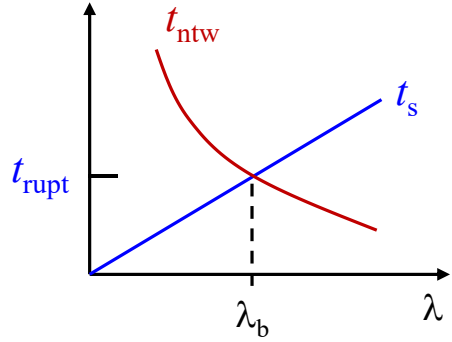

**Figure 1** A schematic illustrating a criterion for elastomeric rupture at stretch rate $\dot{\lambda}$: The straight line represents the elapsed time of stretching, which increases with stretch ratio $\lambda$. The curved line represents the network lifetime ($t_{ntw}$), an intrinsic material property that decreases with the degree of network stretching. Rupture occurs at the intersection when the stretch time $t_s$ elapses to $t_{rupt}$ of Eq. (1), i.e., increasing to the network's lifetime ($t_{ntw}$).

Delayed rupture has also been demonstrated to take place after an incubation time $t_{de\text{-}rupt}$ following a step stretch. The pioneering study[2] of Smith and Stedry revealed $t_{del\text{-}rupt}(T, \lambda_{ss})$ to explicitly depend on temperature T and the degree of step stretching (ratio $\lambda_{ss}$). This phenomenon has been treated in terms of the prevailing framework based polymer viscoelasticity framework.[14] Our analysis and observation contradicts this understanding. We suggest that $t_{del\text{-}rupt}$ reflects the network lifetime, i.e., $t_{del\text{-}rupt}(T, \lambda_{ss}) = t_{ntw}(T, \lambda_{ss})$ of Eq. (2). It follows from the KTBD upon equating Eq. (1) with Eq. (2) that $t_{del\text{-}rupt}(T, \lambda_{ss}) = t_{rupt}(T, \lambda_b)$ at $\lambda_b = \lambda_{ss}$.

One objective of the present study is to offer a fundamentally different interpretation of the nature of delayed rupture, in contrast to that proposed by Smith and Stedry[2] that "the time consumed (for elastomers) in reaching the critical condition for bond decomposition" is far greater than the time required to bond decomposition[17], an assertion[15] originated from Bueche and Halpin. Separating the various timescales, we carried *in situ* birefringence measurements to contradict this flawed paradigm and suggest that delayed rupture in elastomers is not governed by chain dynamics but results from chain scission – an activated process[21] through which network rupture takes time to materialize.

Moreover, in the elastic stretching limit,[23] $TTE_{ntw}$ governs rupture: Rupture will occur at the same value of $\lambda_b$ at two different temperatures when two (non-Hencky) stretch rates $\dot{\lambda}_1$ and $\dot{\lambda}_2$ are chosen to be reciprocal of the lifetime $t_{ntw}$ at the two respective temperatures. According to Eq. (1) and $t_{rupt} = t_{ntw}$, the condition is $\dot{\lambda}_1 t_{ntw}(\lambda_b, T_1) = \dot{\lambda}_2 t_{ntw}(\lambda_b, T_2)$.

In this work, we perform continuous and step stretching of several elastomeric systems along with *in situ* birefringence to investigate rupture and delayed rupture at various temperatures, involving different stretch rates and different magnitudes of step stretching (characterized by $\lambda_{ss}$). Quantitative characterizations allow us to show (a) how lifetime $t_{ntw}(\lambda, T)$ changes with the level of stretching ($\lambda$) and temperature (T) and (b) faster stretching at higher temperatures is equivalent to slower stretching at lower temperatures, i.e., there exists $TTE_{ntw}$, involving $t_{ntw}$ as the pertinent timescale. In conclusion, (1) delayed rupture may be understood in terms of chain scission instead of polymer viscoelasticity, (2) network lifetime varies with temperature in a manner distinctly different from the WLF temperature dependence, and (3) $TTE_{ntw}$ emerges to show that faster stretching to produce rupture at higher temperature is equivalent to slower stretching at a lower temperature.

## 2. Experimental
### 2.1 Sample preparation

The materials used in this study were a commercial acrylate-based elastomer, VHB 4910 (3M product), two crosslinked styrene butadiene rubber (SBR0.1phr and SBR0.03phr), a crosslinked poly(methyl acrylate) (×PMA), and a polyamide elastomer.

The SBR elastomer (SBR0.03phr and SBR0.1phr) were prepared by peroxide-initiated crosslinking. A masterbatch was created by dissolving 100 g of SBR and 0.03 g and 0.1 g of dicumyl peroxide (DCP) crosslinker for SBR0.03phr and for SBR0.1 phr respectively into 400 mL of an appropriate solvent (e.g., toluene). The mixture was stirred for 72 hours to ensure homogeneity. Following mixing, the solvent was removed by drying the solution in a vacuum oven at 70 °C for 24 hours. The resulting rubber compound was then cured into sheets using a hot press at 150 C for 20 minutes. Finally, tensile test specimens were cut from the cured sheets using a home-made dogbone-shaped die cutter with a gauge section measuring 9 mm in length, 1.56 mm in width, and an average final thickness of 1.3 mm. The dogbone-shaped specimens are placed between the clamps at a distance of $L_0$ = 17 mm (cf. inset of Figure 4a).

For the synthesis of the crosslinked poly (methyl acrylate) (PMAx) elastomer, a precursor solution was first prepared. Methyl acrylate (110 mmol, 100 equiv) was combined with an equal volume of chloroform. Butanediol diacrylate (1.1 mmol, 1 equiv) as a crosslinker and Irgacure 819 (0.11 mmol, 0.1 equiv) as a photoinitiator were then dissolved into the mixture. The solution was deoxygenated via a 20-minute nitrogen purge. Under a nitrogen atmosphere, the solution was transferred to a glass-silicone-glass mold (120 × 120 × 1.4 mm). Curing was achieved by irradiating the sample with 365 nm UV light for 1 hour. The resulting film was removed and purified by immersion in toluene for 24 hours, during which the solvent was refreshed three times to remove the sol fraction. Finally, the film was deswollen in methanol, air-dried for 1 hour, and dried under high vacuum at 50 °C for 24 hours. From these finished sheets, tensile test specimens were die cut into a dogbone geometry. The specimens have a gauge section measuring 9.53 mm in length and 3.18 mm in width, an average final thickness of 1.3 mm. The dogbone shaped specimens are placed between the clamps at a distance of $L_0 = 27$ mm (cf. inset of Figure 4b). VHB was cut into such a dogbone shape (cf. inset of Figure 4c).

Synthesis of polyamide elastomer. The diallyl *m*-phthalamide monomer was synthesized as previously reported.[27] To prepare one dog-bone specimen for the fracture analysis, 0.49 g of TEMPIC (0.9 mmol) and 0.34 g of EBMP (1.4 mmol) were pre-dissolved in a scintillation vial using the speed-mixer for 1 minute at 3500 rpm. To this dithiol and trithiol mixture, the synthesized diallyl m-phthalamide monomer (2.7 mmol) and 15 mg of TPO-L was added. This was followed by gently heating to melt and mix the contents of the reaction mixture. The resin was then poured into the pre-heated silicon mold at 70 °C and subjected to UV-curing using a dogbone shaped mold with a gauge section measuring 9.53 mm in length and 3.18 mm in width, an average final thickness of 1.3 mm with the clamping distance of 27 mm at 405 nm at 70 °C for 3 minutes.

2.2 *Methods*

Mechanical testing of the VHB, SBR and ×PMA samples was conducted using an Instron 5969 universal testing machine equipped with a temperature controlled environmental chamber. Samples were mounted using two clamps to perform continuous and stepwise tensile stretching. For delayed rupture experiments, video of the sample deformation was recorded using a 4K camera (Mokose) fitted with a 5-100 mm zoom lens (Arducam).

Continuous uniaxial tensile tests were performed on the SBR samples to investigate the interplay between temperature and strain rate on fracture properties. The experiments were conducted at four different temperatures: room temperature (RT, ca. 25 °C), 50 °C, 70 °C, and 80 °C. At each temperature, a range of crosshead speeds was selected to achieve specific nominal strain rates ($V/L_0$), allowing for the comparison of rupture on different timescales. The specific stretch rate, $V/L_0$, investigated at each temperature were: RT and 80 °C: 0.0005, 0.005, 0.05, and 0.5 s$^{-1}$, 50 °C: 0.0005, 0.005, 0.1, and 0.5 s$^{-1}$ and 70 °C: 0.001, 0.014, and 0.85 s$^{-1}$. This experimental matrix was designed to identify the strain rate required to achieve a targeted range of rupture each temperature, enabling a systematic demonstration of the new time-temperature equivalence. Results are presented in Figures 8a-b.

**3. Results and discussion**

3.1 *In situ* birefringence to elucidate nature of rupture and delayed rupture

We first carried out *in situ* birefringence measurements to show that SBR0.1phr exhibits the same degree of average chain elongation independent of stretch rate used to perform continuous stretching. Figure 2a shows the stress response to be independent of stretch rate. At the same stretch ratio, the birefringence is also the same, as shown in Movies 1 and 2 in Supporting Information (SI). Figure 2b contains four identical-looking birefringence images at the same strain of $\lambda - 1 = 0.37$. Such results remove Bueche's confusion related to his creep tests[15] that "the strength of a rubber should be much greater under a fast than a slow test since in the former case the network chains are not able to elongate very far during the period of the test": In displacement-controlled stretching, chain elongation is affine-like so that birefringence resulting from chain orientation is the same independent of stretching time (stretch rate) required to produce the same nominal strain.

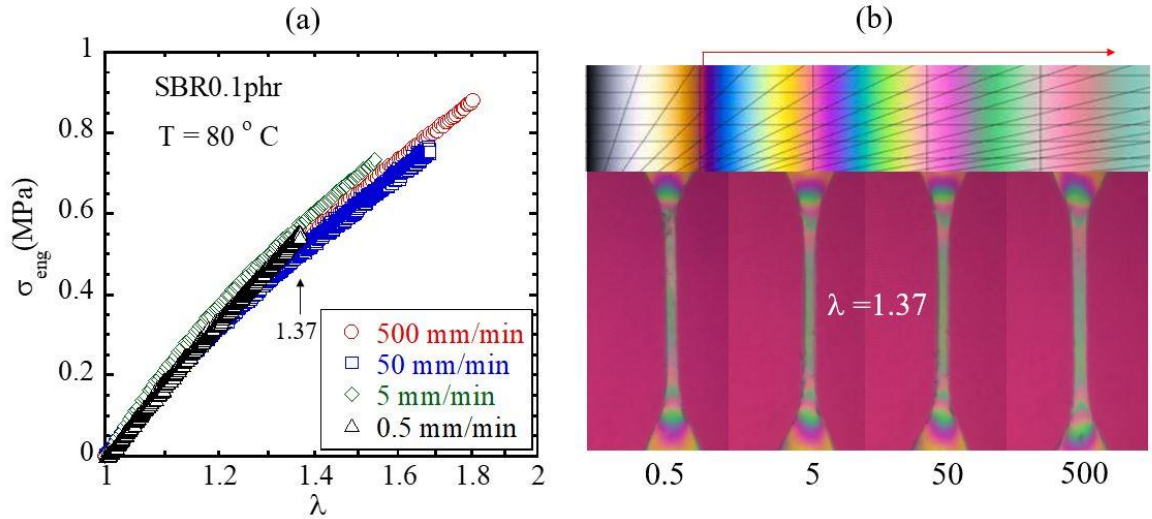

**Figure 2** (a) Engineering stress versus stretch ratio λ for SBR0.1phr with $L_0$=17 mm at 80 °C, stretched with crosshead speed V = 0.5, 5, 50, and 500 mm/min respectively. (b) Birefringence images at λ = 1.37 from all four continuous stretching tests – 1.37 is $λ_b$ for V = 0.5 mm/mm, as indicated by the last triangle in (a). According to the Michel-Levy chart, specimens reached the fourth green at λ = 1.37 in presence of a retardance plate that produces the first red as background.

Motivated by KTBD, we then perform continuous stretching at V = 5 mm/min and observe rupture at ca. $λ_b$ = 1.53. Subsequently, we carry out a step stretch to $λ_{ss}$ = 1.5 ~ $λ_b$ using a high rate, e.g., V = 200 mm/min. Both continuous and step stretching are carried out along with *in situ* birefringence observations in the form of Movies 3 and 4. Figure 3a shows the stress vs. strain curves of these two experiments, confirming the preceding conclusion that stress and chain elongation (shown in Figure 3b) are both independent of stretch rate. Replotting Figure 3a we show in Figure 3c that elapsed time $t_{rupt}$, given by eq 1, at rupture in continuous stretching is of the same magnitude as the incubation time $t_{del-rupt}$ for delayed rupture. Since $t_{del-rupt}$ increases rapidly with decreasing $λ_{ss}$, we can readily separate the polymer relaxation time τ from the incubation time. When $t_{del-rupt} \gg τ$, delayed rupture cannot be explained using the previous paradigm[14] of polymer viscoelasticity-energy dissipation; during the incubation period there is no ongoing stretching and there is little change in the birefringence. The concealed internal clock that dictates when delayed rupture occurs is the network lifetime $t_{ntw}$.

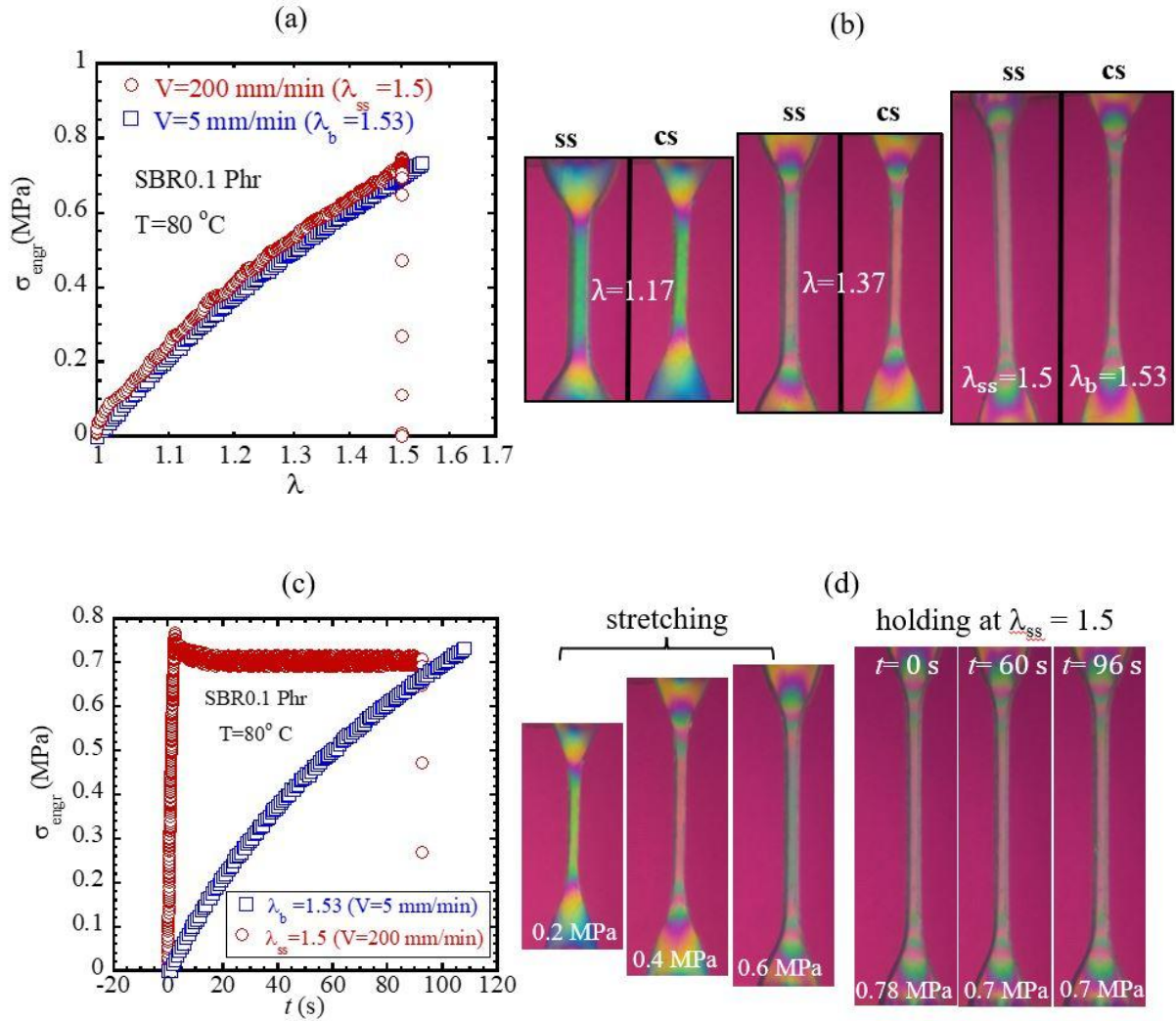

**Figure 3** (a) Uniaxial stretching of SBR0.1phr ($L_0$=17 mm) at 80 °C under respective continuous (with V = 5 mm/min in circles till rupture at $\lambda_b$ = 1.53) and step (with V = 200 mm/min to hold at $\lambda_{ss}$ = 1.5 in squares) stretching conditions. (b) Corresponding birefringence images, comparing chain elongation at common stretch ratios of $\lambda$ = 1.17, 1.37 and ca. 1.5. (c) replot of (a) in time domain. (d) Birefringence images at various stages of step stretch to $\lambda_{ss}$ = 1.5 as well as the last three nearly identical images during the stress relaxation.

Since SBR0.1phr is an elastomer, the level of chain elongation remains essentially unchanged after step stretch, as confirmed by *in situ* birefringence observations in Figure 3d. This result contradicts the expectation of Smith and Stedry[2] that originated from the claim of Bueche[15] and Halpin that elastomers take most of time to creep to the critical strain and chain elongation for bond decomposition and "the time required for decomposition is insignificant"[17] in comparison. It also contradicts any explanation based on energy dissipation or polymer viscoelasticity[14] since there is no ongoing stretching to produce any viscous flow – remarkably Kinloch did not cite this seminal work of Smith and Stedry in his book. Thus, for the first time, we assert that delayed rupture occurs not because step stretch produces chain elongation that could grow after step stretch but because bond dissociation as activated events takes time to produce a path of chain scission in the network, leading to rupture across the width of a macroscopic specimen. Specifically, we interpret the incubation time $t_{\text{del-rupt}}$ for delayed rupture as network's lifetime $t_{\text{ntw}}(\lambda_{\text{ss}}, T)$. In support of the KTBD, $t_{\text{del-rupt}}$ is on the order of $t_{\text{rupt}}$ so that we may draw the conclusion that $t_{\text{del-rupt}}(\lambda_{\text{ss}}, T) = t_{\text{ntw}} \sim t_{\text{rupt}}(\lambda_b, T)$, as shown in Figure 1.

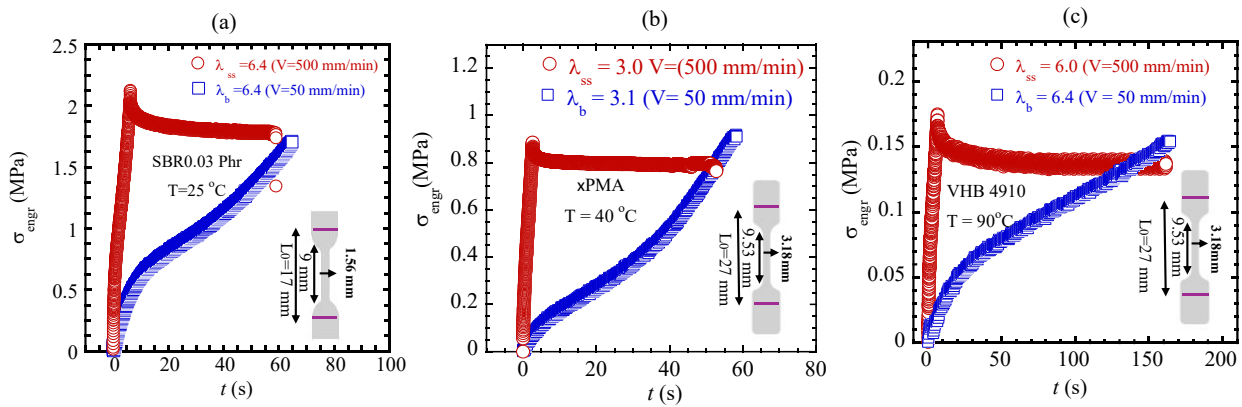

**Figure 4** Engineering stress vs. time under respective continuous and step stretching for (a) SBR0.03phr ($L_0$=17 mm) at room temperature, (b) ×PMA ($L_0$=27 mm) at 40 °C, and (c) VHB ($L_0$=27 mm) at 90 °C, where $\lambda_b$ characterizes the rupture strain and $\lambda_{\text{ss}}$ designates the magnitude of step stretch.

Behavior of SBR0.1phr shown in Figure 3c-d appears universal. In the limit of elastic stretching till rupture and delayed rupture, other elastomers also show such behavior in Figures 4a-c, based on SBR0.03phr at 25 °C, ×PMA at 40 °C and VHB at 90 °C. We will summarize the comparison of timescales involved rupture and delayed rupture below in Section 3.4.

3.2 Pertinent timescale revealed by delayed rupture: temperature dependence

Delayed rupture after step stretching explicitly reveals the hidden internal clock in elastomers. By carrying out step stretching at different temperatures, we measure how this new timescale depends on temperature. Specifically, the incubation time taken for a step-stretched specimen to undergo sudden rupture, $t_{del-rupt}$, reflects the network's lifetime $t_{ntw}$, as given by eq 2. Figure 5a shows that $t_{del-rupt}$ is approximately Arrhenius, i.e., exponentially varying with 1/T. More importantly, the inset figure shows that this temperature dependence systematically deviates from the WLF dependence (solid line). The deviation can be either positive or negative, with $t_{del-rupt}$ showing either stronger or weaker temperature dependence than that of the WLF shift factor $a_T$. For example, unlike VHB, vulcanized SBR exhibits positive deviation, as shown in Figure 5b, consistent with the first study[2] of Smith and Stedry whose Figure 8 also showed[26] a stronger temperature dependence than that prescribed by the WLF factor $a_T$. Since $t_{del-rupt}$ reflects the elastomer's lifetime $t_{ntw}$ of eq 2, the deviation of $t_{del-rupt}$ from the WLF temperature dependence strongly suggests that the internal clock governing elastomeric failure is not polymer relaxation time.

Figure 5b also revealed an interesting feature for $\lambda_{ss}$ = 4. The last data point at room temperature appears to show upward deviation from other data points that form a straight line, exhibiting Arrhenius dependence. The off-Arrhenius behavior is not unique to SBR. For verification, ×PMA is step-stretched at different values of $\lambda_{ss}$. In addition to the negative departure from the WLF temperature dependence, we observe $t_{\text{del-rupt}}$ to show a notable deviation from Arrhenius behavior at lowest test temperatures, as shown in Figure 5c. To confirm this trend, a fourth polyamide based elastomer [27] (a thermoset at elevated temperatures) is studied in step-stretch tests and shown in Figure 5d to reveal a similar upward deviation. This deviation only occurs at lower temperatures, more detailed investigation is beyond the scope of the present study.

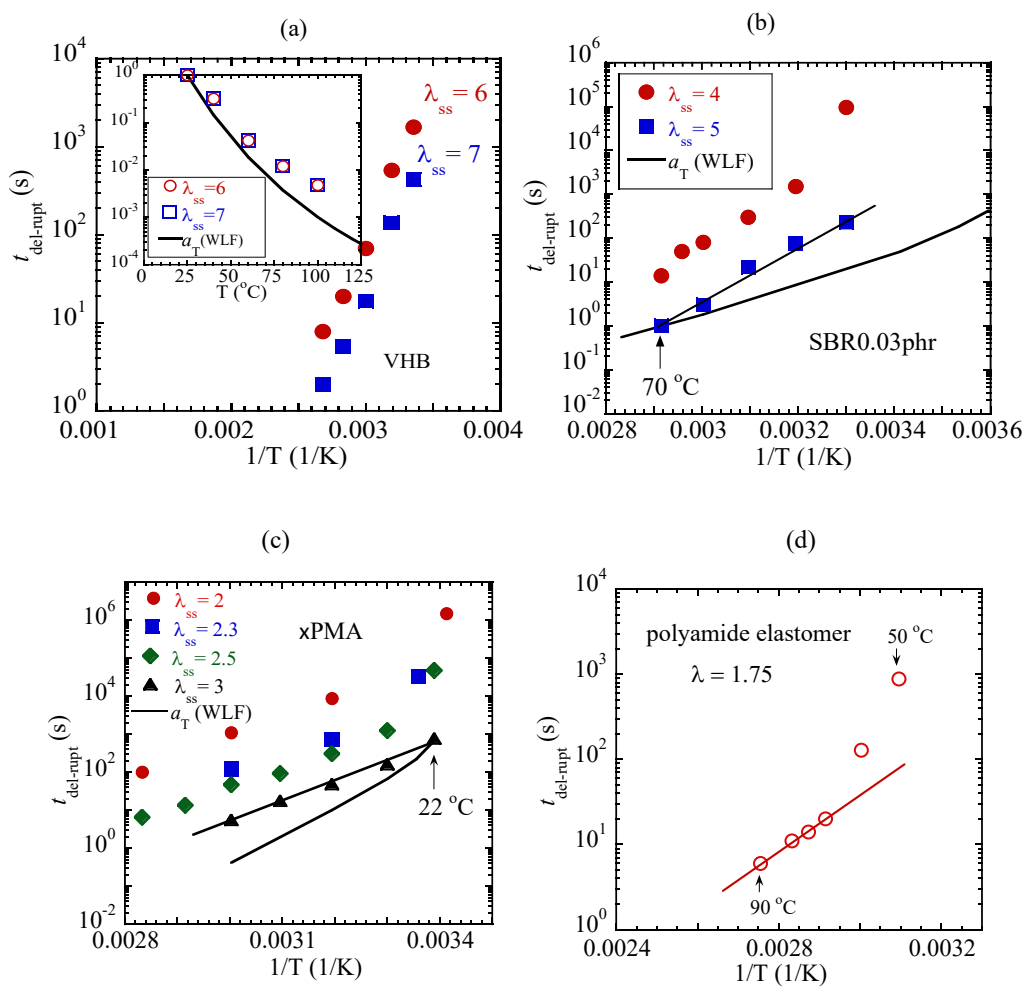

**Figure 5** Dependence of network lifetime $t_{\text{ntw}}$ on temperature, taken as the incubation time $t_{\text{del-rupt}}$ for delayed rupture of four different elastomers after step stretch using cross head speed of 500 mm/min: (a) VHB ($L_0$=27 mm) held at $\lambda_{ss}$ = 6 and 7 respectively, where the inset figure directly compares the temperature dependence of WLF shift factor $a_T$ with that of $t_{\text{del-rupt}}$, (b) SBR0.03phr ($L_0$=17 mm) held at $\lambda_{ss}$ = 4 and 5 respectively, where the smooth curve is WLF shift

factor, (c) ×PMA ($L_0$=27 mm) held at $\lambda_{ss}$ = 2, 2.3, 2.5, and 3 respectively, where the smooth curve is WLF shift factor, (d) polyamide elastomer ($L_0$=27 mm) held at $\lambda_{ss}$ = 1.75.

Having conducted continuous stretching tests[1] Smith and Stedry anticipated[2] delayed rupture perhaps because they thought that their SBR just needed more time after a fast step stretching for chains to achieve critical molecular strain for chain decomposition. For example, referring to Figure 3a, it took ca. 100 s for rupture (squares) to occur, and step stretch only took 2.55 s. Therefore, according to Smith and Stedry, we should wait for about 97 s for delayed rupture to take place. These 97 seconds were understood by Bueche, Halpin, Smith and most others to be required for molecular strain to be established for chain decomposition. However, our birefringence measurements in Figure 3b indicate that the molecular strain is already achieved in 2.55 s. According to our KTBD, the stretched network has a lifetime of ca. 100 s at this stretch ratio of 1.5, beyond which bond dissociation destroyed the network in the form of rupture.

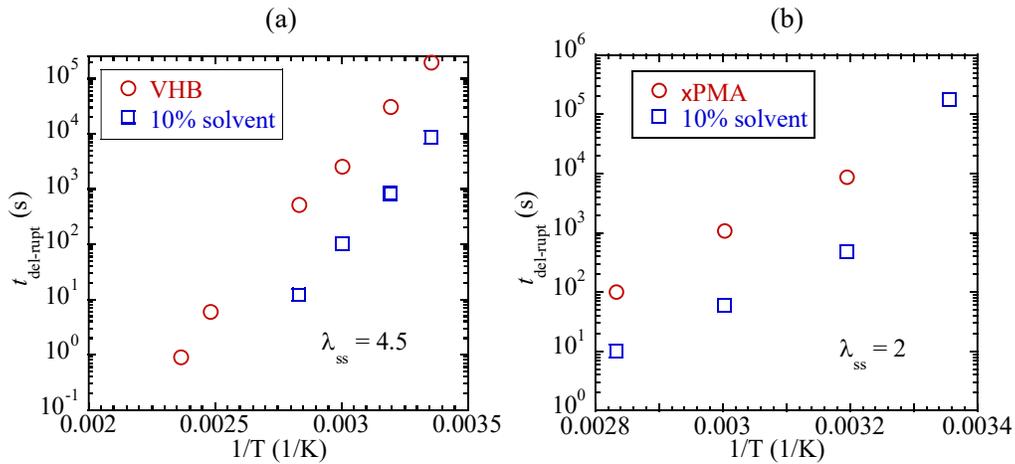

**Figure 6** Solvent effect on delayed rupture: (a) incubation time $t_{del-rupt}$ for delayed rupture of (a) of VHB ($L_0$=27 mm) at various temperatures, (b) PMA($L_0$=27 mm), both shorter by a decade by the solvents (squares).

### 3.3 Solvent effect on network lifetime

Solvent effects have been suggested to minimize polymer viscoelasticity and reduce energy dissipation.[20, 28] Rubber swelling by solvent is used to access the fatigue threshold.[29, 30] In the KTBD description of the structural origin of elastomeric failure,[21] we suggest that solvent may accelerate the network lifetime. To demonstrate the solvent effect, we incorporate less than 10% of dibutyl adipate into VHB and 10% of dimethyl sulfoxide into ×PMA so that their elastic responses and stress relaxation behavior are little affected, as shown in Figures SI.1a-b in Supporting Information (SI). Figures 6a-b indicate that the incubation time $t_{\text{del-rupt}}$ is shortened by a decade in the presence of solvent for both VHB and ×PMA, where circles represent $t_{\text{del-rupt}}$ of solvent-free elastomers.

### 3.4 Comparison of two timescales

Stretching with different rates $\dot{\lambda}$ produces rupture on timescales ($t_{\text{rupt}}$) approximately inversely proportional to $\dot{\lambda}$ per eq 1. In other words, given the explanation of Figure 1, characteristics of rupture, i.e., $t_{\text{rupt}}$ along with $\lambda_b$ can be used to reveal the network lifetime over a range as wide as that of the applied rate. Figure 2a, along with Figures SI.2a-c (VHB, SBR, PMA) shows that with increasing stretch rate rupture occurs at measurably higher $\lambda_b$. Since strain ($\lambda_b-$1) at rupture only logarithmically increases with $\dot{\lambda}$, applied rate $\dot{\lambda}$ approximately discloses the network lifetime per eq 1.

To measure network lifetime $t_{\text{ntw}}$ we consider step-stretching to different magnitudes, characterized by stretch ratio $\lambda_{ss}$, using the highest available stretch speed. For a range of $\lambda_{ss}$, we measure $t_{\text{del-rupt}}(\lambda_{ss})$ as shown in circles in Figure 7a-c based on three different elastomers. In squares, Figure 7a-c shows $t_{\text{rupt}}$ as a function of $\lambda_b$ from corresponding continuous stretching tests with stretch rates $\dot{\lambda} = V/L_0 = 0.0005, 0.005, 0.05$, and $0.5$ s$^{-1}$ for SBR0.1phr, $\dot{\lambda} = 0.0031, 0.031$ and $0.31$ s$^{-1}$ for VHB, and $\dot{\lambda} = V/L_0$ and $\dot{\lambda} = V/L_0 = 0.00031, 0.0031, 0.031$ and $0.31$ s$^{-1}$ for ×PMA, where the ellipses mark the three separate cases where the raw data were respectively presented in Figure 2c and Figure 4b-c.

The overlap of circles ($t_{del-rupt}$) and squares ($t_{rupt}$) over several decades reveals that $t_{del-rupt}(T, \lambda_{ss}) = t_{rupt}(T, \lambda_b)$ at $\lambda_{ss} = \lambda_b$. Since $t_{del-rupt}$ may be taken as $t_{ntw}$, the overlapping in Figure 7a-c confirms the underlying physics illustrated in Figure 1: $t_{rupt} = t_{ntw}$, i.e., rupture occurs when elastomers' lifetime becomes comparable to the stretch time. In other words, in the elastic stretching limit rupture tests also reveal the lifetime. We conclude that the network lifetime $t_{ntw}$ is dictated by the degree of stretching, i.e., $\lambda_{ss}$ for delayed rupture in step stretch and by $\lambda_b$ for rupture in continuous stretch.

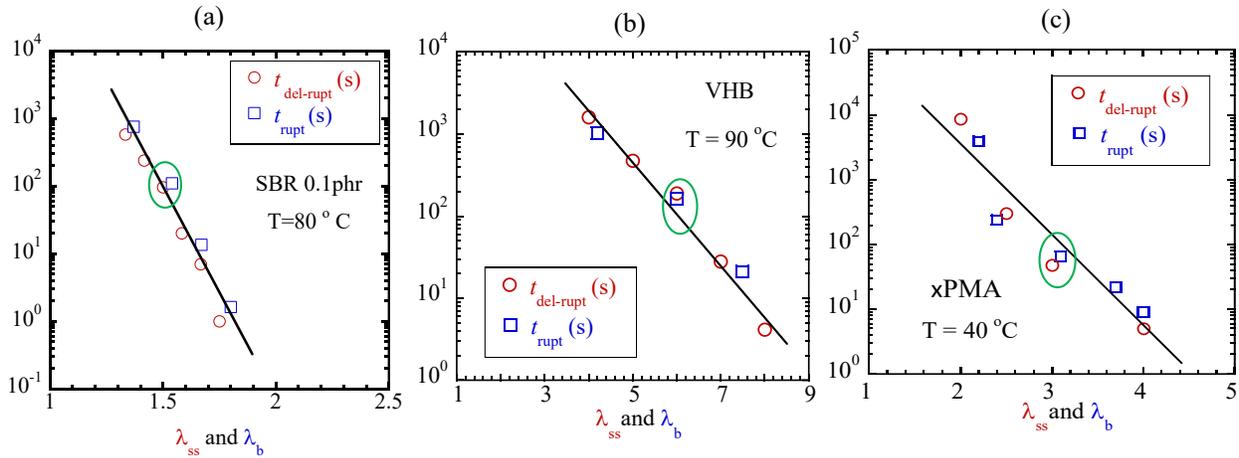

**Figure 7** Two timescales, $t_{del-rupt}$ and $t_{rupt}$ obtained from step and continuous stretching involving various step-stretch ratio $\lambda_{ss}$ and rupture stretch ratio $\lambda_b$, based on (a) SBR0.1phr ($L_0$=17 mm) at 80 °C, (b) VHB ($L_0$=27 mm) at 90 °C, and (c) ×PMA ($L_0$=27 mm) at 40 °C.

## 3.5 Time-temperature equivalence in rupture

Like the polymer relaxation time, the network lifetime is also temperature dependent (cf. eq 2), as shown in Figure 5a-d. In rheology of uncrosslinked melts,[12] faster deformation at higher temperatures is equivalent to slower deformation at lower temperatures. Analogously, since changing temperature changes the network lifetime which dictates when rupture occurs, there is new time-temperature equivalence. Faster stretching at higher temperature is equivalent to slower stretching at lower temperature. This new $TTE_{ntw}$ is based on the temperature dependence of lifetime $t_{ntw}(T)$ and has nothing to do with the familiar $TTE_{ve}$ in polymer rheology that involves the chain relaxation time.

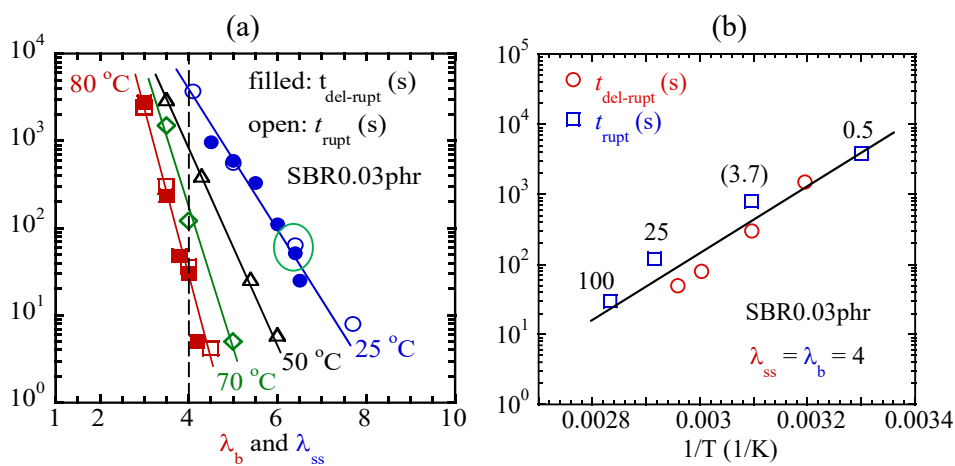

**Figure 8** (a) Two timescales, $t_{del-rupt}$ and $t_{rupt}$ obtained from step and continuous stretching involving various step-stretch ratio $\lambda_{ss}$ and rupture stretch ratio $\lambda_b$, based on SBR0.03phr ($L_0$= 17 mm) at four temperatures of 25, 50, 70 and 80 °C. (b) New time-temperature equivalence (squares), confirmed by the temperature dependence of network lifetime (circles). The same rupture at $\lambda_b$ = 4 are produced at four different temperatures using four different stretch rates, indicated, where the value of 3.7 mm/min at 50 °C is read from (a) from the extrapolating straight line.

Because of $TTE_{ntw}$, we can identify a pair of values for stretch rate $\dot\lambda$ and temperature T that produces the same rupture strain, i.e., the same $\lambda_b$. For this purpose, we carry out both continuous and step stretching of SBR0.03phr at four temperatures to show four sets of data in Figure 8a, including the pair at 25 °C in the ellipse whose raw data were already presented in Figure 4a. Raw data at other temperatures can be found in Figure SI.4a-c in SI. At 80 °C, $\lambda_{ss}$ and $\lambda_b$ falls in a range from 3 to 4.5 whereas the range is from 4 to 7.7 at 25 °C. The change in $\lambda_{ss}$ and $\lambda_b$ at different temperatures reflects the temperature dependence of $t_{ntw}$: at lower temperatures more stretching is required to produce the same lifetime.

Reading Figure 8a vertically along the dashed line, for example, at $\lambda_b = 4$, we show in Figure 8b that these rates produce rupture at different times given by $t_{rupt}$ in squares as a function of temperature. The data of $t_{del-rupt}$ from delayed rupture in circles are from Figure 5b. The overlap has similar meaning to that of Figure 7a-c: $t_{rupt}$ of eq 1 reflects the network lifetime $t_{ntw}$. Figure 8a shows how the two timescales change with the degree of network stretching whereas Figure 8b shows how they change with temperature.

In the temperature range from 25 to 80 °C, varying stretch rate from $5 \times 10^{-4}$ to 0.5 s$^{-1}$ can equivalently produce the same rupture, i.e., rupture at the same $\sigma_b$ and $\lambda_b$. In other words, Figure 8b, obtained from Figure 8a, reveals that rupture at $\lambda_b = 4$ in a range of temperatures from 25 to 80 °C involves a change in $t_{ntw} = t_{rupt}$ (squares) of by ca. two decades, produced by different stretch rates from 0.1 (80 °C) to $5 \times 10^{-4}$ (25 °C) s$^{-1}$. Such a demonstration of TTE$_{ntw}$ has added benefit: it automatically reveals the temperature dependence of $t_{ntw} = t_{rupt}$, where the equality follows from the approximate collapses of squares and circles.

## 4. Conclusion

Guided by the central idea that bond dissociation in backbones of most stretched network strands controls characteristics of elastomeric rupture, we carried out continuous and step stretching under a wide range of stretch rates, aiming to verify the foundation of the KTBD. Two key ingredients or pillars of the KTBD are (a) network lifetime $t_{ntw}$ exponentially decreasing with network stretching and (b) rupture taking place when $t_{ntw}$ decreases during continuous stretching till $t_{ntw} = t_{rupt}$, which is the elapsed (stretch) time at rupture. As described in the last paragraph of Section 3.2, the rheological and rheo-optical observations show that elastomers, extended in displacement-controlled stretching, consume little time to reach the critical condition for bond decomposition and spend most of the time waiting for bond dissociation to result in rupture.

The present study demonstrates the following key results. First, step stretch tests at various temperatures show how the incubation time for delayed rupture, $t_{del-rupt}$, depends (1) on temperature (cf. Figure 5a-d) and (2) on imposed strain in an exponential manner (cf. Figure 7a-c). Second, continuous stretch tests with various stretch rates at different temperatures disclose how stretch rates define the time window, through which elastomers' lifetime $t_{ntw}$ emerges as the elapsed time $t_{rupt}$ at rupture as a function of temperature and degree of stretching. Third, comparison between $t_{del-rupt}$ and $t_{rupt}$ as a function of imposed strain $\lambda_{ss}$ or $\lambda_b$ (cf. Figure 7a-c) proves that rupture is indeed a moment when elastomeric lifetime $t_{nwt}$, which is measured by $t_{del-rupt}$, equals the elapsed time $t_{rupt}$. Fourth, there is new time-temperature equivalence (TTE$_{ntw}$) based on the lifetime $t_{ntw}$. Under elastic stretching conditions where this new TTE$_{ntw}$ holds, fast stretching at high temperatures is equivalent to slow stretching at low temperatures. Similarly, at a given temperature, faster stretching produces rupture on a shorter timescale at larger strain because only higher strain makes the elastomer's lifetime shorter.

**Acknowledgments**

This work is supported, in part, by the Polymers program through Special Creativity Extension of the US National Science Foundation grant DMR-2210184. We thank Zehao Fan for small-amplitude oscillatory shear measurements of WLF shift factors of VHB, SBR and PMA. JPW acknowledges NSF funding CHE-2204079, and CW's team at Utah is supported, in part, by the startup fund from Price College of Engineering, University of Utah.

**CrediT authorship contribution statement**

**Asal Y. Siavoshani:** Investigation, Visualization, Methodology, Data curation, Formal analysis. Writing of Supporting Information. Cheng Liang: Data curation. Ming-Chi Wang and Aanchal Jaisingh: Syntheses of PMA and polyamide thermoset respectively. Chen Wang: Supervision of AJ. Junpeng Wang: Supervision of MCW. **Shi-Qing Wang:** Conceptualization. Investigation. Writing – original draft. Writing – review & editing. Validation. Formal analysis. Supervision of AYS and CL.